\documentclass{article}

\usepackage[nonatbib, preprint]{neurips_2025}

\usepackage[utf8]{inputenc} 
\usepackage[T1]{fontenc}    
\usepackage{hyperref}       
\usepackage{url}            
\usepackage{booktabs}       
\usepackage{amsfonts}       
\usepackage{nicefrac}       
\usepackage{microtype}      
\usepackage[table]{xcolor}         
\usepackage{wrapfig}

\usepackage{graphicx}
\usepackage{amsmath}
\usepackage{amssymb}
\usepackage[ruled,vlined]{algorithm2e} 
\usepackage{mathtools}
\usepackage{dashrule}
\usepackage{xargs}
\usepackage{caption}
\usepackage{lipsum}
\usepackage{comment}
\usepackage{multirow}
\usepackage{caption}
\usepackage{multibib}
\usepackage{siunitx}
\usepackage{enumitem}
\usepackage[most]{tcolorbox}
\tcbset{
  promptbox/.style={
    colback=gray!10,         
    colframe=gray!50,        
    boxrule=0.5pt,           
    arc=2mm,                 
    left=5pt,                
    right=5pt,               
    top=5pt,
    bottom=5pt,
    enhanced,
    sharp corners=south,     
    title={#1},              
    fonttitle=\bfseries,     
  }
}

\let\titleold\title
\renewcommand{\title}[1]{\titleold{#1}\newcommand{\thetitle}{#1}}
\def\maketitlesupplementary
   {
   \newpage
       {
        \centering
        \Large
        \textbf{\thetitle}\\
        \vspace{0.5em}Supplementary Material \\
        \vspace{1.0em}
       } 
   }

\newcites{supp}{Supplementary}

\graphicspath{{figs/}}

\title{Generative Panoramic Image Stitching}

\author{%
    Mathieu Tuli\textsuperscript{*} \\ 
    LG Electronics \\
    \texttt{tuli.mathieu@gmail.com} \\
    \And
    Kaveh Kamali\textsuperscript{*}  \\
    LG Electronics \\ 
    \texttt{kaveh.kamali@lge.com} \\
    \And
    David B. Lindell \\
    LG Electronics \\ 
    University of Toronto \\ 
    \texttt{lindell@cs.toronto.edu} \\
}

\definecolor{lightgray}{RGB}{200, 200, 200}
\definecolor{tabfirst}{RGB}{255, 215, 0}
\definecolor{tabsecond}{RGB}{210, 210, 255}
\definecolor{tabthird}{RGB}{240, 240, 240}
\definecolor{link}{RGB}{240, 50, 150}

\newcommand{\tile}{\mathbf{b}}
\newcommand{\numtiles}{B}

\newcommand{\img}{\mathbf{x}}
\newcommand{\imgref}[1]{\mathbf{x}_\text{ref}^{(#1)}}
\newcommand{\imgpanos}[1]{\mathbf{x}_\text{pano}^{(#1)}}
\newcommand{\imgpano}{\mathbf{x}_\text{pano}}
\newcommand{\imgpos}{\mathbf{x}_{\gamma}}
\newcommand{\numref}{N}
\newcommand{\mask}{\mathbf{m}}
\newcommand{\validmask}{\mathbf{m}_\text{valid}}

\newcommand{\latentimg}{\mathbf{z}}

\newcommand{\sdencoder}{\mathcal{E}_\text{SD}}
\newcommand{\ctxencoder}{\mathcal{E}_\text{ctx}}

\newcommand{\cond}{\mathcal{C}}
\newcommand{\condctx}{\mathbf{c}_\text{ctx}}

\newcommand{\noise}{\boldsymbol{\epsilon}}
\newcommand{\model}{\Psi}

\begin{document}

\maketitle

\begin{center}
\vspace{-4.5em}
\hspace{-13em}
\small{\textnormal{\textsuperscript{*}joint first authors}}\\
\vspace{2.4em}
\end{center}

\begin{figure*}[h]
  \vspace{-2.5em}
  \includegraphics[width=\textwidth]{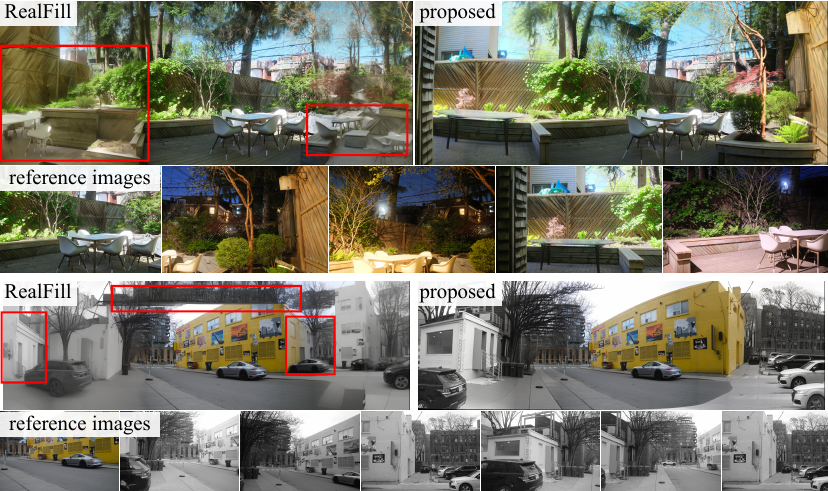}
  \vspace{-1.5em}
  \caption{
 We introduce a generative method for panoramic image stitching from multiple casually captured reference images that exhibit strong parallax, lighting variation, and style differences. Our approach fine-tunes an inpainting diffusion model to match the content and layout of the reference images. After fine-tuning, we outpaint one reference image (e.g., the leftmost reference view shown here) to create a seamless panorama that incorporates information from the other views. Unlike prior methods such as RealFill~\cite{tang2024realfill}, which produces artifacts when outpainting large scene regions (red boxes), our method more accurately preserves scene structure and spatial composition.}
  \label{fig:teaser}
\end{figure*}

\begin{abstract}
We introduce the task of generative panoramic image stitching, which aims to synthesize seamless panoramas that are faithful to the content of multiple reference images containing parallax effects and strong variations in lighting, camera capture settings, or style. In this challenging setting, traditional image stitching pipelines fail, producing outputs with ghosting and other artifacts. While recent generative models are capable of outpainting content consistent with multiple reference images, they fail when tasked with synthesizing large, coherent regions of a panorama. To address these limitations, we propose a method that fine-tunes a diffusion-based inpainting model to preserve a scene's content and layout based on multiple reference images. Once fine-tuned, the model outpaints a full panorama from a single reference image, producing a seamless and visually coherent result that faithfully integrates content from all reference images. Our approach significantly outperforms baselines for this task in terms of image quality and the consistency of image structure and scene layout when evaluated on captured datasets.

\end{abstract}

\section{Introduction}
Creating a coherent visual representation from multiple input images is a long-standing problem in computer vision~\cite{szeliski2007image,szeliski1997creating},
and many techniques have been proposed to combine multiple images from different perspectives to synthesize panoramas~\cite{brown2007automatic}, multi-perspective images~\cite{agarwala2006photographing,peleg2000mosaicing,seitz2003multiperspective}, or photo montages~\cite{agarwala2004interactive,nomura2007scene}. 
More recently, image generation models make it possible to render or outpaint new image content based on one or more input images~\cite{ruiz2023dreambooth,tang2024realfill}.
Inspired by methods for panorama synthesis and recent image generation techniques, we propose to address the task of \textit{generative panoramic image stitching}---i.e., we seek to generate seamless panoramas that are faithful to the content of multiple reference images captured from different viewpoints with strong variations in lighting or style (Figure~\ref{fig:teaser}).

\begin{wrapfigure}[20]{r}{2.1in}
    \vspace{-2.5em}
    \includegraphics[width=2.1in]{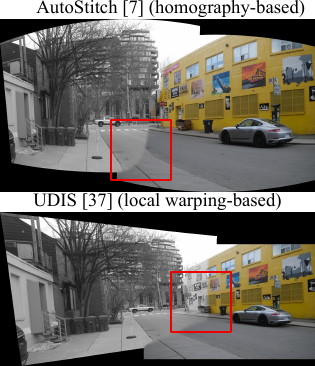}
    \caption{Conventional panoramic image stitching methods~\cite{brown2007automatic,nie2023parallax} fail to account for strong parallax effects or variations in lighting or style.}
    \label{fig:failures}
\end{wrapfigure}


A standard approach for panoramic image stitching involves detecting feature correspondences and estimating geometric transformations between input images~\cite{brown2007automatic}. 
Then, the input images are warped based on the estimated transformation and blended together into a panorama~\cite{burt1983multiresolution,perez2003poisson}.
Conventionally, these techniques use a homography to relate input images, which assumes that there is no parallax (i.e., no translation between captured viewpoints)~\cite{szeliski2007image}. 
Violating this assumption results in artifacts, such as ghosting~\cite{eden2006seamless}, as shown in Figure~\ref{fig:failures} (top).
Hence, a significant amount of effort has been devoted to improving robustness to viewpoint changes, e.g., by optimizing local warping operations~\cite{chang2014shape,gao2011constructing,lee2020warping,li2017parallax,liao2019single,lin2015adaptive,lin2011smoothly,zaragoza2013projective,zhang2014parallax}, by using graph cuts to minimize seams between blended images~\cite{eden2006seamless,gao2013seam,zhang2014parallax}, or optimizing neural networks~\cite{nie2021unsupervised,nie2023parallax}, but completely avoiding artifacts is challenging when images are captured from significantly different positions.  
Further, standard techniques for image stitching assume that camera acquisition settings and illumination conditions are roughly constant across input images; while image blending can help to mitigate small variations in camera gain, exposure, white balance, or scene illumination~\cite{brown2007automatic,burt1983multiresolution,perez2003poisson} it fails to handle strong variations in the lighting or style of input images (Figure~\ref{fig:failures}, bottom).

A separate line of work seeks to create panoramic images via image synthesis. 
For example, using generative models, recent approaches synthesize panoramas from a text prompt~\cite{bar2023multidiffusion,frolov2025spotdiffusion,lee2023syncdiffusion,ye2024diffpano} or inpaint masked regions of an input panorama~\cite{wu2024panodiffusion}; however, these methods do not handle the stitching of reference images with overlapping fields of view and significant parallax effects.
The recent work of Tang et al.~\cite{tang2024realfill} uses a pre-trained image generation model for reference-guided inpainting, which is close to our task.
Specifically, they fine-tune an image diffusion model to inpaint a set of casually captured reference images from different viewpoints and lighting conditions.
After fine-tuning, the model can be used to outpaint an existing image in a way that is consistent with the content of the reference images and robust to parallax or lighting variations.
However, we find that attempting to use this approach for panoramic image stitching fails, as outpainting large missing regions results in artifacts and scene layouts that are not faithful to the input reference images (see Figure~\ref{fig:teaser}).  

Here, we address limitations of conventional methods for panoramic image stitching as well as more recent, reference-driven outpainting techniques~\cite{tang2024realfill}.
Given a set of casually captured reference images, we first compute a coarse alignment of the images via conventional feature matching and homography estimation~\cite{brown2007automatic}, resulting in a set of warped images and their approximate locations on an initial panorama.
To correct artifacts in this initial panorama---such as those caused by parallax or lighting inconsistencies---we fine-tune~\cite{ruiz2023dreambooth} a large, pre-trained inpainting diffusion model~\cite{stablediffusioninpaint} to solve a position-aware inpainting task. 
Specifically, we fine-tune the model to inpaint and outpaint each warped input image while conditioning on positional encodings that reflect the image’s location within the panorama.
Once fine-tuned, the model is used to iteratively outpaint the panorama from a single reference image, resulting in a seamless composite that integrates content from all reference views as shown in  Figure~\ref{fig:teaser}.\\\\

In summary, we make the following contributions.
\begin{itemize}[leftmargin=*,noitemsep,topsep=0pt]
    \item We propose the task of \textit{generative panoramic image stitching}, which seeks to generate panoramas that are faithful to a set of reference images containing significant parallax effects and variations in illumination or style.
    \item We address this task with a method that estimates the coarse layout of the reference images within a panorama and then fine-tunes a diffusion model to generate a seamless output panorama via position-aware outpainting.
    \item We evaluate our approach on a dataset of captured images and show state-of-the-art results for this task compared to baselines based on reference-driven image outpainting and image stitching. 
\end{itemize}

\section{Related Work}
Our work also connects to other methods for learning-based image stitching, multi-perspective rendering, 3D reconstruction, and reference-driven outpainting. 

\paragraph{Learning-based image stitching.}
While conventional image stitching pipelines typically use feature-based homography estimation~\cite{szeliski2007image}, other approaches directly regress a homography using a neural network~\cite{detone2016deep,le2020deep,nguyen2018unsupervised} or learned features~\cite{zhang2020content}, which can improve performance for dynamic scenes or images with limited texture. 
Nie et al.~\cite{nie2021unsupervised,nie2023parallax} introduce a two-stage procedure for image stitching that first predicts a homography between two input images using a neural network and then warps the resulting image using a transformer or thin-plate splines to reduce stitching artifacts. 
Our procedure uses a similar two-stage approach, but we leverage a standard feature-based approach for the initial alignment~\cite{brown2007automatic}, which we find generalizes well to our captured in-the-wild images. 
Then, instead of directly warping the input images, we leverage generative priors and position-aware inpainting and outpainting to synthesize a seamless panorama.
As such, our approach scales to handle multiple input images,
and we avoid stitching artifacts due to parallax or lighting variations because the output panorama is synthesized by the generative model rather than produced by warping the input images.

\paragraph{Multi-perspective rendering and 3D reconstruction.}
It is also possible to synthesize panoramas using image-based rendering~\cite{agarwala2006photographing,anderson2016jump,liu2009content,nomura2007scene,rav2008minimal}.
Given a sufficiently densely captured set of input images, one can directly capture or estimate the desired set of light rays used to assemble an output panorama or multi-perspective image~\cite{anderson2016jump,bergen1991plenoptic,levoy1996light,richardt2013megastereo}.
Alternatively, one can reconstruct a 3D representation of the scene and render novel views from any desired viewpoint~\cite{rav2008minimal,wood2000surface,debevec1996modeling,buehler2001unstructured,mildenhall2021nerf}.
Still, these techniques cannot be easily applied to our proposed task, where only a few images are provided as input, camera poses are unknown, and the images have inconsistencies, e.g., due to variations in camera capture settings, color palette, or lighting. 

\paragraph{Reference-driven image editing.}
Rather than directly stitching the input images, our approach generates a panorama by outpainting one of the input views using content from the others. This design is motivated by prior work on reference-driven inpainting. For instance, Yang et al.\cite{yang2023paint} inpaint masked regions of an image using objects from a reference image depicting a different scene. Zhou et al.\cite{zhou2021transfill} extend this idea to multiple images from the same scene. Most similar to our method, Tang et al.\cite{tang2024realfill} fine-tune a diffusion model for reference-guided outpainting; however, their method does not incorporate scene layout information and fails in the context of panorama synthesis (see Fig.\ref{fig:teaser}).

\section{Generative Panoramic Image Stitching}

\begin{figure*}[t]
  \includegraphics[width=\textwidth]{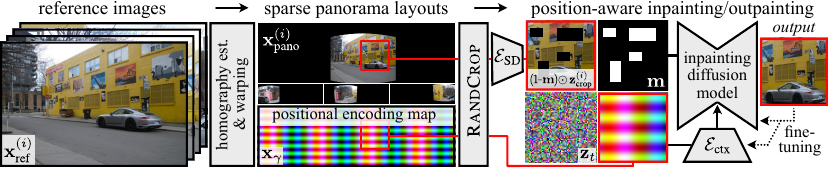}
  \vspace{-1.5em}
  \caption{Method overview. Given a set of reference images $\{\imgref{i}\}_{i=1}^{N}$, we generate sparse panorama layouts $\{\imgpanos{i}\}_{i=1}^{N}$ by detecting features~\cite{lowe2004distinctive}, estimating homographies, and warping each reference image to its location in a sparse panorama containing only that image. We then fine-tune a pre-trained inpainting diffusion model for a position-aware inpainting/outpainting task. During training, random crops are taken from the sparse panoramas and a positional encoding map $\imgpos$. 
Each panorama crop is processed using an encoder $\sdencoder$, and we we multiply the resulting latent image $\latentimg^{(i)}_\text{crop}$ with a random binary mask $(1-\mask)$. 
We process the crop of $\imgpos$ with an encoder $\ctxencoder$ and use the result to condition the diffusion model. 
The other inputs --- the masked version of $\latentimg_\text{crop}^{(i)}$, the mask $\mask$, and the noisy latent image $\latentimg_t$ --- are concatenated together and passed as input to the model. After fine-tuning, we generate seamless panoramas by outpainting one of the initial sparse panoramas.}
  \label{fig:method}
  \vspace{-1em}
\end{figure*}

We introduce our approach by first providing a brief background on latent diffusion models. 
Then, we describe our method for generative panoramic image stitching based on (1) initial panorama layout estimation via homography estimation and warping, (2) fine-tuning a diffusion model for position-aware panorama inpainting and outpainting, and (3) generating a seamless panorama via iterative outpainting. 
An overview of the approach is shown in Figure~\ref{fig:method}.

\subsection{Preliminaries: Latent Diffusion Models}
Latent diffusion models~\cite{sohl2015deep} are based on a forward and reverse process
that either gradually introduces noise or removes it from a latent image $\latentimg_0$ --- i.e., an image encoded into the latent space of an image autoencoder. 
Latent images are typically lower in resolution than conventional images and so operating in the latent space yields improvements in computation and memory~\cite{rombach2022high}.


More specifically, latent diffusion models use a Markovian forward process to iteratively transform the latent image $\latentimg_0$ into standard Gaussian noise $\latentimg_T \sim \mathcal{N}(\mathbf{0}, \mathbf{I})$ over $T$ time steps. 
The intermediate noisy images $\latentimg_t$ produced during this process are defined as~\cite{ho2020denoising}
\begin{equation}
\latentimg_t = \sqrt{\alpha_t} \latentimg_0 + \sqrt{1-\alpha_t}\noise,
\label{eqn:noise}
\end{equation}
where $\noise \sim \mathcal{N}(\mathbf{0}, \mathbf{I})$ and the set of values
$\{\alpha_t\}_{t=1}^T$ defines a fixed noise schedule such that increasing $t$ corresponds to adding more noise. 
In turn, the reverse process estimates $\latentimg_{t-1}$ by gradually denoising from $\latentimg_T \sim \mathcal{N}(0, \mathbf{I})$. 
The clean latent image $\latentimg_0$ is generated through a reverse diffusion process by iteratively predicting the noise $\noise$ at each time step using a neural network $\model$. 
Then, applying a decoder network converts the latent image into a conventional image $\img$. 

In the \textit{conditional} reverse process, the network is trained to predict
the noise by minimizing the loss
\begin{equation}
\mathcal{L} = \mathbb{E}_{\latentimg, t, \noise} \parallel \model(\latentimg_t, t, \cond) - \noise \parallel_2^2.
\end{equation}
Here, the network is given a conditioning signal $\cond$---e.g., a text prompt or a masked image for inpainting. 
At inference time, $\model$ is sampled to remove noise from $\latentimg_T$ and iteratively estimate $\latentimg_{t-1}$ until the clean latent image $\latentimg_0$ is recovered.

\subsection{Panorama Layout Estimation \& Positional Encoding}
\label{sec:layout}

Given a set of $\numref$ input reference images \( \{\imgref{i}\}_{i=1}^{N} \), where \( \imgref{i} \in \mathbb{R}_{+}^{H_\text{ref} \times W_\text{ref} \times 3} \), we aim to generate a panorama \( \imgpano \in \mathbb{R}^{H_\text{pano} \times W_\text{pano} \times 3} \) via latent diffusion that seamlessly stitches together scene content from the reference views and outpaints uncaptured scene regions. 

The first step in this procedure involves producing an initial panorama layout via homography estimation and warping. 
We adapt the procedure of Brown et al.~\cite{brown2007automatic} to detect feature correspondences between the input images, estimate homographies, and warp each image into cylindrical coordinates.
The result of this procedure is a set of sparse panoramas $\{\imgpanos{i}\}_{i=1}^{N}$, $\imgpanos{i}\in\mathbb{R}_{+}^{H_\text{pano}\times W_\text{pano}\times 3}$, which each contains a single warped reference image (see Figure~\ref{fig:method}). 

We also associate the panorama with a positional encoding map $\imgpos$~\cite{mildenhall2021nerf}. 
The map is computed using a function $\gamma(p) = [\cos(\pi f_1 p), \sin(\pi f_1 p),  \ldots, \cos(\pi f_F p), \sin(\pi f_F p)]^T$, where $\{f_i\}_{i=1}^{F}$ are the encoding frequencies, and the function $\gamma(\cdot)$ is applied to each vertical and horizontal pixel coordinate $p$. 
Encoding each pixel coordinate results in $\imgpos\in\mathbb{R}^{H_\text{pano} \times W_\text{pano} \times 4F}$, where $F$ is the number of positional encoding frequencies. Additional details are provided in Supp.\ Section~\ref{sec:supp-data-prep}. 

\subsection{Fine-tuning for Position-aware Inpainting and Outpainting}
We use the set of panoramas $\{\imgpanos{i}\}_{i=1}^{N}$ and the positional encoding map $\imgpos$ to fine-tune an inpainting diffusion model for position-aware inpainting and outpainting.


\paragraph{Architecture.}
Our approach adapts a pre-trained inpainting diffusion model $\model(\latentimg_t, t, \cond)$ (we use Stable Diffusion 2.1~\cite{stablediffusioninpaint}).
The model is conditioned on the input
\begin{equation}
\cond = \{\mask, (1-\mask) \odot \latentimg_\text{crop}^{(i)}, \condctx \},
\end{equation}
where $\mask$ is a randomly generated binary mask to be inpainted or outpainted, $\odot$ indicates Hadamard product, and $\latentimg_\text{crop}^{(i)}$ is a randomly cropped region of $\imgpanos{i}$ that we encode into the latent space using the Stable Diffusion encoder $\sdencoder$, or $\latentimg_\text{crop}^{(i)} = \mathcal{E}_{\text{SD}}(\textsc{RandCrop}(\imgpanos{i}))$.
The context embedding tensor $\condctx$ is produced as $\condctx = \ctxencoder(\textsc{RandCrop}(\imgpos))$, where we apply the same random crop to the positional encoding map $\imgpos$ as for $\imgpanos{i}$. 
We process the cropped version of $\imgpos$ using $\ctxencoder$, a small three-layer convolutional encoder with a linear layer (see Supp.\ Section~\ref{sec:supp-posenc}).

While the context embedding tensor $\condctx$ is used by the pre-trained model for text conditioning, our approach repurposes it to encode the positional information, and we provide the tensor as input to the cross-attention layers of the network.
The other conditioning signals (i.e., $\mask$ and $(1-\mask)\odot \latentimg_{\text{crop}}^{(i)}$) are concatenated with the noisy latent image $\latentimg_t$ and passed as input to the diffusion model. 
A more detailed description of the architecture is provided in Supp.\ Section~\ref{sec:supp-details}.

\paragraph{Optimization.}
We fine-tune the network $\model$ to minimize the loss function
%
\begin{equation}
    \mathcal{L} = \mathbb{E}_{\noise, \latentimg_\text{crop}^{(i)}, i, t, \mask} \left[ \left\| \validmask\odot\left(\model(\latentimg_{\text{crop},t}^{(i)}, t, \cond ) - \noise\right) \right\|_2^2 \right],
\label{eqn:loss}
\end{equation}
where $\validmask$ is a binary mask that restricts the loss to regions of $\latentimg_\text{crop}^{(i)}$ that correspond to non-empty areas in the cropped sparse panorama $\imgpanos{i}$.
Hence, we fine-tune the model to minimize the difference between the noise it predicts and the noise added to $\latentimg^{(i)}_\text{crop}$, where $\latentimg^{(i)}_{\text{crop},t} = \sqrt{\alpha_t}\latentimg^{(i)}_{\text{crop}} + \sqrt{1-\alpha_t}\noise$ (as described in Equation~\ref{eqn:noise}).

\begin{wrapfigure}[14]{r}{2.3in}
\vspace{-2.5em}
\includegraphics[width=2.3in]{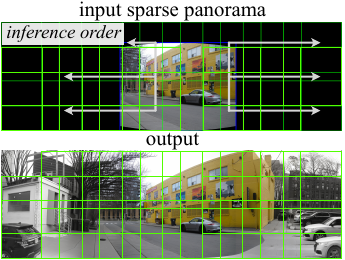}
\caption{Panorama generation. We use the fine-tuned model to iteratively outpaint each tile (green grid) of the panorama, in order of distance from the center (white arrows), to create a seamless result.}
\label{fig:inference}
\end{wrapfigure}

We use low-rank adaptation (LoRA)~\cite{hu2022lora} to optimize the model's self-attention layers and preserve the capabilities of the Stable Diffusion model's pre-trained weights. 
The cross-attention layers undergo full-parameter fine-tuning to better adapt to the positional encoding information provided by $\condctx$. 
Last, we initialize and optimize all parameters of the context encoder $\ctxencoder$.

\subsection{Panorama Generation}
After fine-tuning, we generate a seamless panorama $\img_\text{pano}$ by outpainting one of the initial sparse panoramas $\imgpanos{i}$. 
The main challenge in this step is that the resolution of the panorama is much larger than the nominal resolution for which the inpainting diffusion model is trained---so we cannot generate the entire panorama in a single inference pass. 
Instead, we sequentially denoise tiles of the panorama to generate the final output as depicted in Figure~\ref{fig:inference}. 
We apply the sequential denoising procedure to the sparse panorama containing a centered warped reference image --- this is an arbitrary image to which we register the other reference images during the initial layout estimation process (Section~\ref{sec:layout}).

Specifically, we generate an evenly spaced grid of overlapping image tiles or boxes $\{\tile^{(i)}\}_{i=1}^{\numtiles}$ across the panorama, where $\tile^{(i)}=\{x^{(i)}, y^{(i)}, H, W\}$ gives the pixel coordinates of the corner of the tile and the height and width of the tile. 
In practice, we use $20\%$ overlap between tiles, and we set $H=W=512$. 
After positioning the tiles, if some tiles extend beyond the extent of the panorama, the overlap is reduced until all tiles fit within the panorama in both the vertical and horizontal dimensions.
For each tile in the grid, we run the full reverse diffusion process using the DDPM sampler~\cite{ho2020denoising} to inpaint/outpaint the missing regions of the tile. 
Inpainting/outpainting masks are feathered and composited with the current state of the generated panorama $\img_\text{pano}$.
Different than training, where we randomly sample the mask values $\mask$, during inference we set the $\mask$ values to indicate which regions of each input tile have not yet been generated. The tiles are denoised in order of increasing distance from their centroids to that of the warped reference image.
We summarize this procedure in Algorithm~\ref{alg:inference}.


\begin{wrapfigure}[19]{R}{0.6\textwidth}
    \vspace{-5.5em}
\begin{minipage}{0.58\textwidth}
    \begin{algorithm}[H]
\caption{Panorama Generation}
\label{alg:inference}
\KwIn{$\imgpano^{(0)}$ \hfill \raisebox{0.1em}{\scalebox{0.7}{$\blacktriangleleft$}} Input sparse panorama \\
    \hspace{2.8em} $\imgpos$ \hspace{1em} \hfill \raisebox{0.1em}{\scalebox{0.7}{$\blacktriangleleft$}} Positional encoding map \\
    \hspace{2.7em}  $\{\tile^{(i)}\}_{i=1}^\numtiles$ \hfill \raisebox{0.1em}{\scalebox{0.7}{$\blacktriangleleft$}} Tiles covering panorama\\
}
\vspace{0.2em}
\KwOut{Panorama $\img_{\text{pano}}$ with all tiles denoised}
\vspace{0.2em}
\tcp{Order tiles based on distance}
\small $[\tile^{(1)}, \tile^{(2)}, \ldots, \tile^{(\numtiles)}]\leftarrow \textsc{OrderByDist}(\{\tile^{(i)}\}_{i=1}^{\numtiles}) $

\For{$i = 1$ \KwTo $\numtiles$}{
    \tcp{Initialize noisy latent by adding noise to reference latent}
    $\latentimg_{\tile^{(i)},T} \leftarrow \sdencoder(\img^{(0)}_\text{pano}[\tile^{(i)}]) + \noise,\quad \boldsymbol{\epsilon} \sim \mathcal{N}(0, \sigma^2 I)$\\
$\condctx \leftarrow \ctxencoder(\imgpos[\tile^{(i)}])$ \\
$\mask \leftarrow \textsc{GetRegionToInpaint}(\latentimg_{\tile^{(i)},T})$\\
\vspace{0.2em}
    \For{$t = T$ \KwTo $1$}{
        \tcp{Denoise tile}
        $\latentimg_{\tile^{(i)}, t-1} \leftarrow \textsc{Denoise}(\latentimg_{\tile^{(i)}, t},
        \mask, (1-\mask) \odot \latentimg_{\tile^{(i)}, t-1}, \condctx)$
    }
    \tcp{Update output with denoised tile using latent decoder $\mathcal{D}_\text{SD}$}
    $\img_\text{pano}[\tile^{(i)}] \leftarrow \textsc{Composite}(\img_\text{pano}[\tile^{(i)}], \mathcal{D}_\text{SD}(\latentimg_{\tile^{(i)}, 0}))$
}

\Return $\img_\text{pano}$
\end{algorithm}
\end{minipage}
\end{wrapfigure}

\subsection{Implementation Details}
\label{sec:impl}


\paragraph{Masking and augmentation.} 
Inpainting masks are synthesized with randomly generated patterns following Tang et al.~\cite{tang2024realfill}. 
We also introduce an augmentation scheme which perturbs the location of the warped images in the sparse panorams with a random similarity transformation. We find that this helps to avoid seams from appearing in the final output panoramas at the boundary locations of the warped images.

\paragraph{Training and inference.} 
We apply LoRA to the Stable Diffusion model’s self-attention layers and fully fine-tune the cross-attention layers and $\ctxencoder$, using AdamW with learning rates of $\num{1e-4}$ (LoRA), $\num{3e-4}$ (cross-attention), and $\num{8e-4}$ ($\ctxencoder$). Training runs for 4,000 iterations with batch size 32 and takes 4.5 hours on 2$\times$A100 GPUs. At inference, a $1000 \times 3000$ panorama typically takes 1 minute to generate on a single RTX 2080 Ti. We use classifier-free guidance~\cite{ho2021classifier} with $\condctx = 0$ and a guidance scale of 1.5.

\paragraph{Correspondence-based seed selection.} 
We employ a correspondence-based seed selection process~\cite{tang2024realfill} to identify generated panoramas whose layout matches the result of feature-based image registration~\cite{brown2007automatic}.
Specifically, we generate ten panoramas with different random seeds and take our output to be the panorama with the most feature matches (computed with LoFTR~\cite{sun2021loftr}) compared to the reference.
Please see Supp.\ Section~\ref{sec:supp-details} for additional implementation details.

\section{Experiments}
\label{sec:exp}

\paragraph{Dataset.}
We collect two image datasets of eight scenes each, with several images captured for each scene.
One dataset consists of \textit{tripod-captured} images collected by rotating a camera on a tripod, and a set of \textit{casually captured} images from different scene viewpoints using a handheld camera (Fujifilm X100 VI).
In the casually captured dataset, the distance between viewpoints varies by up to one to two meters, and we also introduce other challenging variations, such as capturing images of the same scene with varying illumination conditions, camera white balance, or image color palette.

The tripod-captured dataset, with minimal parallax, aligns with assumptions of standard stitching methods and is used to compute a reference panorama for evaluating image quality. The casually captured dataset tests robustness to parallax, illumination, and style variations. A detailed description of the captured scenes and the number of captured images for each scene is provided in Supp.\ Section~\ref{sec:supp-data}.

To facilitate comparison across output panoramas, we include one tripod-captured image within the set of casually captured images. We configure our method and all baselines so that this shared image is placed at the center of the output panorama, ensuring a consistent layout across output panoramas from both sets of images.

\begin{figure}[t]
    \begin{minipage}{\textwidth}
    \includegraphics[width=\textwidth]{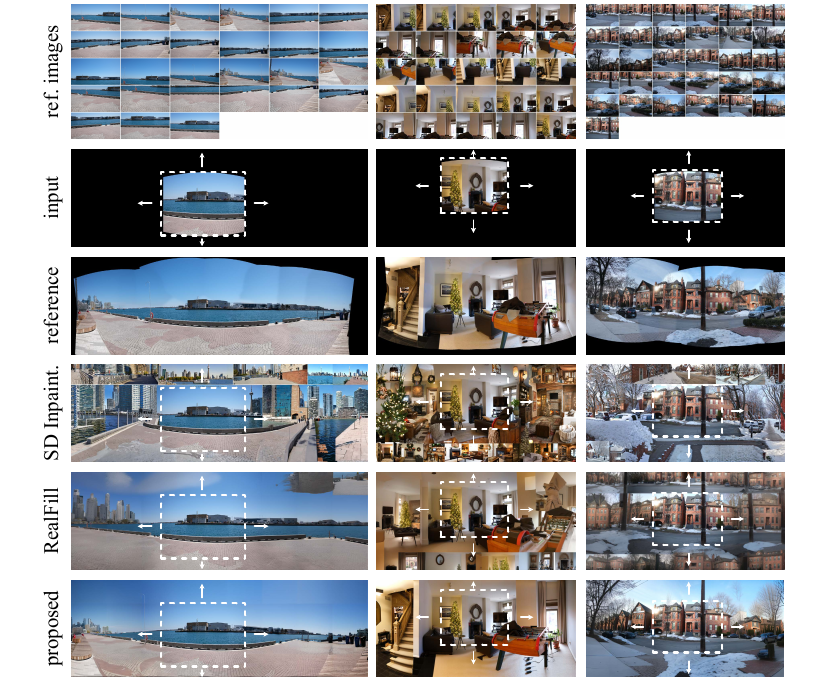}
    \vspace{-1em}
    \captionof{figure}{Qualitative results on the tripod-captured dataset. We find that our approach produces panoramas that are more consistent with the layout and content of the reference panorama than baseline approaches based on inpainting/outpainting.}
    \label{fig:tripod}
    \end{minipage}

    \begin{minipage}{\textwidth}
    \small
    \centering
    \setlength{\tabcolsep}{1pt}
    \resizebox{\textwidth}{!}{%
    \begin{tabular}{lcccccccc}
         \toprule
         Method &  PSNR (dB) $\uparrow$ & SSIM  $\uparrow$& LPIPS $\downarrow$ & DreamSim $\downarrow$ & DINO $\uparrow$ & CLIP $\uparrow$ & LoFTR (L2 Distance) $\downarrow$ & LoFTR (Matching) $\uparrow$\\ \toprule
         \toprule
         \input{sec/tables/table1}\\
         \bottomrule
 \end{tabular}}
        \captionof{table}{Quantitative assessment of generative panoramic image stitching on the tripod-captured image dataset. We outpaint a single warped reference image (see Figure~\ref{fig:tripod}), and compare the generated result to a reference panorama produced using AutoStitch~\cite{brown2007automatic}. Our approach generates panoramas that are most faithful to the reference images. Standard deviations are reported in parentheses.}
    \label{tab:tripod}
\end{minipage}
\vspace{-2em}
\end{figure}

\vspace{-.5em}
\paragraph{Baselines.}
We compare our approach to multiple baselines, starting with the conventional image stitching method of Brown and Lowe~\cite{brown2007automatic} (AutoStitch), which uses feature matching, homography estimation, warping, and blending.
We chose this baseline because (1) it informs our own panorama layout estimation step, and (2) we found, through empirical evaluation, that it was more robust than other methods for parallax-tolerant stitching.
In particular, we found AutoStitch’s bundle adjustment procedure more effective for stitching multiple images than recent methods tailored to pairwise stitching or reliant on pre-trained networks, which failed to generalize to our captured image datasets.

We also compare to the Stable Diffusion 2 inpainting model~\cite{stablediffusioninpaint}, which serves as the backbone of our method. This baseline omits our positional encoding and fine-tuning strategy, but follows the same iterative outpainting procedure.  Additionally, we compare to RealFill~\cite{tang2024realfill}, using their inpainting-based fine-tuning strategy and generating panoramas using our iterative outpainting process. For the Stable Diffusion 2 baseline, we use a guidance scale of $7.5$ during inference, and for RealFill we follow their implementation and do not use guidance.



\vspace{-1em}
\paragraph{Metrics.} 
We evaluate our method using standard image quality metrics, learning-based metrics that assess high-level image structure, and feature-matching-based metrics that assess how well our approach preserves the scene layout.
Specifically, we use standard image quality metrics: peak signal-to-noise ratio, structure similarity~\cite{wang2004image}, and learned perceptual image patch similarity~\cite{zhang2018unreasonable}.  
To evaluate high-level image structure, we use DreamSim~\cite{fu2023dreamsim}, which assesses similarity in semantic content and layout.
We also compute the cosine similarity between the DINO~\cite{caron2021emerging} and CLIP~\cite{radford2021learning} full-image embeddings.
Additionally, we use image feature matches from LoFTR~\cite{sun2021loftr} to assess how well the layout of the output panorama matches a reference. 
We report both the L2 distance between the pixel coordinates of matching features and the number of matched features divided by the total number of features in the reference image (see Supp. Section~\ref{sec:supp-metrics} for more details).

\begin{figure}[t]
    \begin{minipage}{\textwidth}
    \includegraphics[width=\textwidth]{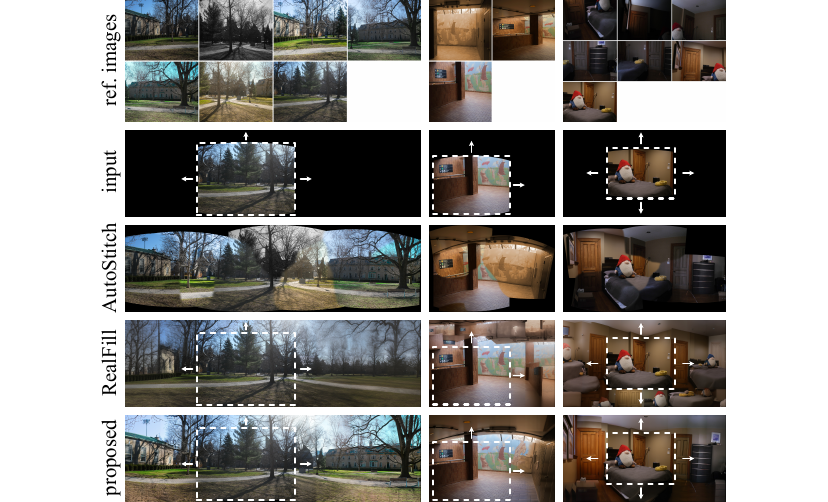}
    \vspace{-1.5em}
    \captionof{figure}{Qualitative results on the casually captured dataset. Even in this challenging scenario, where the input images have strong parallax effects and variations in style, illumination, color palette, or camera capture settings, our approach reconstructs seamless panoramas that preserve the content and layout of the reference.}
    \label{fig:casual}
\end{minipage}
\begin{minipage}{\textwidth}
    \small
    \centering
    \setlength{\tabcolsep}{1pt}
    \resizebox{\textwidth}{!}{%
    \begin{tabular}{lcccccccc}
         \toprule
         Method &  PSNR $\uparrow$ & SSIM  $\uparrow$& LPIPS $\downarrow$ & DreamSim $\downarrow$ & DINO $\uparrow$ & CLIP $\uparrow$ & LoFTR (L2 Distance) $\downarrow$ & LoFTR (Matching) $\uparrow$\\ \toprule
         \input{sec/tables/table2}\\
         \bottomrule
 \end{tabular}}
    \vspace{-0.5em}
    \captionof{table}{Quantitative assessment of generative panoramic image stitching from casually captured images. We compare the generated results to a reference panorama using AutoStitch~\cite{brown2007automatic} on the tripod-captured dataset. Our approach generates panoramas that are close to the reference despite operating on images with parallax and variations in style or lighting.}
    \label{tab:casual}
    \vspace{-2em}
\end{minipage}
\end{figure}

\vspace{-1em}
\paragraph{Qualitative results.}
We show qualitative results on the tripod-captured dataset in Figure~\ref{fig:tripod} and on the casually-captured image datasets in Figures~\ref{fig:teaser} and~\ref{fig:casual}.
For the tripod-captured dataset we observe that the Stable Diffusion inpainting model~\cite{stablediffusioninpaint} produces image content that is locally plausible, but fails to adhere to the layout and content of the actual scene.
RealFill~\cite{tang2024realfill} improves on this result, but tends to repeat scene content from the reference images without respecting the actual scene layout.
Our approach provides a much closer match to the layout provided by the reference panorama while also resolving seams and avoiding ghosting artifacts. 

For the casually captured results in Figure~\ref{fig:casual}, we compare to AutoStitch~\cite{brown2007automatic}, which fails to convincingly blend between the different image regions, resulting in ghosting and other artifacts. 
We see similar artifacts for RealFill as in the tripod-captured dataset, and we find that our approach produces seamless results that are more consistent with the layout and content of the scene.
Additional results for all scenes are included in Supp.\ Section~\ref{sec:supp-results}.

\paragraph{Quantitative results.}
We report quantitative results on the tripod-captured and casually captured image datasets in Tables~\ref{tab:tripod} and Tables~\ref{tab:casual}, respectively.
For the tripod-captured dataset, we construct a reference panorama using the method AutoStitch~\cite{brown2007automatic}, which is well-suited to these images, as they have minimal parallax or variations in illumination.  
We find that the proposed approach generates panoramas that are significantly more consistent with the layout of the reference panorama than the baselines.
This trend is clear from the qualitative results as well as the metrics that assess similarity in high-level image structure (e.g., DreamSim, CLIP) and in layout based on feature matching (LoFTR).  
We note that low-level image quality metrics (e.g., PSNR, SSIM) are perhaps less useful for assessing performance on this task because small variations in layout can produce large changes in the pixel values.
Nevertheless, these metrics follow the same trend as the high-level metrics, and we include them for completeness.

For the casually captured image dataset, we compare the output of our approach and baselines to the same reference panorama as before (i.e., computed with the tripod-captured dataset). For RealFill and our proposed method, we employ correspondence-based seed selection.
Since the set of input images differs from that of the reference panorama, we notice worse performance in the low-level image quality metrics on this dataset. However, our approach still outperforms baselines for most metrics. 
We notice similar trends in the high-level image quality metrics to those of the tripod-captured dataset, which suggests that our approach retains the same layout and structure as the reference despite the significantly more challenging setting.
While Autostitch~\cite{brown2007automatic} performs slightly better than our method on the feature-matching based metrics, it achieves this at a cost of seams and other artifacts because it imperfectly accounts for parallax and variations in capture settings or illumination. 

\vspace{-0.5em}
\paragraph{Ablation study.}
\begin{table}
    \small
    \centering
    \setlength{\tabcolsep}{1pt}
    \resizebox{\textwidth}{!}{%
    \begin{tabular}{lcccccccc}
         \toprule
         Method &  PSNR $\uparrow$ & SSIM  $\uparrow$& LPIPS $\downarrow$ & DreamSim $\downarrow$ & DINO $\uparrow$ & CLIP $\uparrow$ & LoFTR (L2 Distance) $\downarrow$ & LoFTR (Matching) $\uparrow$\\ \toprule
         proposed (w/o perturb) &                      10.83 (1.76) & \cellcolor{tabsecond}0.394 (0.100) &                      0.547 (0.058) &                      0.146 (0.011) & \cellcolor{tabsecond}0.971 (0.014) &  \cellcolor{tabthird}0.920 (0.042) &                      20.90 (3.76) &                      0.099 (0.040) \\
proposed (w/o LoRA)    &  \cellcolor{tabthird}11.00 (1.89) &  \cellcolor{tabfirst}0.420 (0.102) &  \cellcolor{tabthird}0.534 (0.054) &  \cellcolor{tabthird}0.142 (0.019) &  \cellcolor{tabfirst}0.972 (0.012) &  \cellcolor{tabfirst}0.923 (0.032) & \cellcolor{tabsecond}19.37 (5.06) &  \cellcolor{tabthird}0.117 (0.042) \\
proposed (random seed) & \cellcolor{tabsecond}11.24 (2.07) &  \cellcolor{tabthird}0.375 (0.140) & \cellcolor{tabsecond}0.514 (0.076) & \cellcolor{tabsecond}0.139 (0.034) & \cellcolor{tabsecond}0.971 (0.012) & \cellcolor{tabsecond}0.922 (0.034) &  \cellcolor{tabthird}19.95 (7.48) & \cellcolor{tabsecond}0.121 (0.052) \\
\hline
proposed               &  \cellcolor{tabfirst}11.35 (2.15) &                      0.374 (0.143) &  \cellcolor{tabfirst}0.508 (0.076) &  \cellcolor{tabfirst}0.137 (0.033) & \cellcolor{tabsecond}0.971 (0.013) &                      0.917 (0.035) &  \cellcolor{tabfirst}17.97 (5.14) &  \cellcolor{tabfirst}0.130 (0.056)
\\
         \bottomrule
 \end{tabular}}
 \vspace{0.5em}
    \captionof{table}{Ablation study. We evaluate the effects of omitting (1) the similarity transform used to perturb the location of the warped images in the sparse panoramas (w/o perturb.), (2) LoRA and instead using full fine-tuning on the model's self-attention layers (w/o LoRA), and (3) correspondence-based seed selection (random seed). We compare the generated results to a reference panorama produced using AutoStitch~\cite{brown2007automatic} on the tripod-captured dataset.}
    \label{tab:ablation}
    \vspace{-3.0em}
\end{table}

We conduct an ablation study on four scenes from the casually captured dataset (see  Table~\ref{tab:ablation}). 
We evaluate (1) not perturbing the warped image positions (Section~\ref{sec:impl}), (2) replacing LoRA with full fine-tuning of the self-attention layers, and (3) using a single random seed instead of correspondence-based seed selection. 
Without perturbation, the model is less robust to misalignments in the initial layout estimation; full fine-tuning shows no significant advantage over LoRA and is more computationally expensive; correspondence-based seed selection improves the overall image quality and feature similarity.
In Supp.\ Section~\ref{sec:supp-ablations} we provide additional ablations to evaluate the effects of guidance scale, seed selection, tiling strategy, and positional encoding frequencies.

\vspace{-1em}
\section{Discussion}
\label{sec:discussion}

\begin{wrapfigure}[13]{r}{2in}
\vspace{-4.0em}
\includegraphics{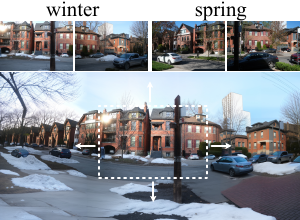}
\vspace{-1.5em}
\caption{Generative stitching result (bottom) with changing scene content in the reference images (top). See Supp.\ Section~\ref{sec:supp-data} for all reference images for this scene.}
\label{fig:winter-spring}
\end{wrapfigure}

Our work overcomes several failure cases associated with conventional panoramic image stitching methods and shows the utility of image generation methods for this low-level computer vision task.
We see multiple promising directions for future work. 
While our method is currently fine-tuned on a single scene, future extensions could train a more general model that can incorporate layout and content from multiple reference images in a feed-forward fashion.
Additionally, we demonstrate how our method can handle input images with large variations in viewpoint, lighting, white balance, and color palette. 
However, strong variations in scene content, such as dynamic scenes with many moving objects, can be challenging to handle with our layout estimation scheme, which leverages conventional feature matching and homography estimation. 
In Fig.~\ref{fig:winter-spring} we show that our method is relatively robust to scene changes, e.g., using images captured in winter and spring, but tackling highly dynamic scenes remains an interesting challenge.


\paragraph{Broader impact.}
\label{sec:broader}
In contrast to conventional image-stitching methods, we use an image generation model that can hallucinate scene content. Hence, our method should be used for applications where the qualitative appearance of an output panorama is more important than the strict pixel-level fidelity.


{
    \small
    \bibliographystyle{plain}
    \bibliography{ref}
}

\clearpage
\setcounter{page}{1}

\maketitlesupplementary

\setcounter{figure}{0}
\setcounter{table}{0}
\setcounter{equation}{0}
\setcounter{section}{0}

\renewcommand{\thesection}{S\arabic{section}}
\renewcommand{\thefigure}{S\arabic{figure}}
\renewcommand{\thetable}{S\arabic{table}} 
\renewcommand{\theequation}{S\arabic{equation}} 



\section{Supplementary Implementation Details}
\label{sec:supp-details}

\subsection{Data Preparation}
\label{sec:supp-data-prep}

We use reference images capturing a scene from multiple viewpoints to build a panorama. Images are aligned via homography-based registration and warped onto a panorama of size \( H_\text{pano} \times W_\text{pano} \) using homography matrices \( \mathbf{H}_i \), computed with feature matching (we use SIFT~\cite{lowe2004distinctive,brown2007automatic}). 

A global positional encoding \( \imgpos \in \mathbb{R}^{H_\text{pano} \times W_\text{pano} \times C} \), with \( C \in \{4, 8, 12\} \), encodes spatial patterns. Coordinates \( x \in [0, W_\text{pano}] \), \( y \in [0, H_\text{pano}] \) are normalized:
\[
p_x = \frac{2x - W_\text{pano}}{W_\text{pano}}, \quad p_y = \frac{2y - H_\text{pano}}{H_\text{pano}}.
\]
Frequencies are:
\[
f_i = \exp\left( \log f_{\text{min}} + \frac{i (\log f_{\text{max}} - \log f_{\text{min}})}{F} \right), \quad i = 0, \ldots, F - 1.
\]
The encoding is:
\[
\imgpos(y, x, c) =
\begin{cases}
\sin(\pi p_x f_{c/2}), & c = 2i, \\
\cos(\pi p_x f_{c/2}), & c = 2i + 1, \\
\sin(\pi p_y f_{(c - 2F)/2}), & c = 2F + 2i, \\
\cos(\pi p_y f_{(c - 2F)/2}), & c = 2F + 2i + 1,
\end{cases}
\]
where \( F = C / 4 \) is the number of frequencies per dimension, \( i = 0, \ldots, F - 1 \), and \( c \) denotes the channel index. In our proposed method, we use  \(f_{\text{min}}=1\), \(f_{\text{max}}=50\) and  \(C=12\).

\subsection{Positional Encoding Processing}
\label{sec:supp-posenc}

A crop of the global positional encoding \( \imgpos[\tile] \in \mathbb{R}^{H_{\text{pano}} \times W_{\text{pano}} \times C} \), where $\tile=\{x, y, W=512, H=512\}$, is transformed into context embeddings \( \condctx \in \mathbb{R}^{B \times 77 \times 1024} \) (where \( B \) is the batch size) through a convolutional processing module. The transformation proceeds as follows:

\begin{align*}
F_1 &= \text{GELU}\left( \text{Conv2D}(\imgpos[\tile],\ \text{kernel}=4,\ \text{stride}=2,\ \text{padding}=1,\ \text{out}=128) \right), \\
F_2 &= \text{GELU}\left( \text{Conv2D}(F_1,\ \text{kernel}=4,\ \text{stride}=2,\ \text{padding}=1,\ \text{out}=128) \right), \\
F_3 &= \text{AdaptiveAvgPool2D}(F_2,\ (7,\ 11)),
\end{align*}

reducing the spatial dimensions from \( 512 \times 512 \) to \( 7 \times 11 \). The feature map is then reshaped and projected to a higher-dimensional space:

\begin{align*}
F_4 &= \text{Reshape}(F_3,\ (B,\ 77,\ 128)), \\
F_5 &= \text{Linear}(F_4,\ \text{out}=1024),
\end{align*}

yielding \( F_5 \in \mathbb{R}^{B \times 77 \times 1024} \). A token-level positional encoding inspired by transformer-based language models~\cite{vaswani2017attention} is added to \( F_5 \). This encoding is precomputed for all 77 token positions and uses sinusoidal functions to encode each token's position in the sequence. For each token index \( p \in \{0, 1, \ldots, 76\} \), its corresponding embedding \( PE(p) \in \mathbb{R}^{1024} \) is defined by:
\[
PE(p, 2i) = \sin\left(p / 10000^{2i / d}\right), \quad PE(p, 2i+1) = \cos\left(p / 10000^{2i / d}\right),
\]
for \( i \in \{0, 1, \ldots, d/2-1\} \), where \( d = 1024 \) is the embedding dimension. The token positional encoding is added to \( F_5 \), i.e., \( F_6 = F_5 + PE \), where \( PE \in \mathbb{R}^{77 \times 1024} \) is broadcast across the batch dimension. Finally, a layer normalization is applied:
\[
\condctx = \text{LayerNorm}(F_6),
\]
producing the final context embeddings \( \condctx \in \mathbb{R}^{B \times 77 \times 1024} \), which are used as input to the pre-trained inpainting diffusion model's cross-attention layers. The entire positional encoding processor network is trained from scratch, allowing it to optimally learn the transformation from spatial positional encodings to meaningful context embeddings for the panorama inpainting and outpainting task.

\subsection{Attention Mechanisms}
\label{sec:supp-attention}

The model employs self-attention and cross-attention to integrate internal features across multiple views and enforce spatial consistency, respectively.

\subsubsection{Self-Attention}
Self-attention operates on the latent feature map \( \latentimg \in \mathbb{R}^{B \times C \times H \times W} \) (e.g., \( C = 320, H = W = 64 \)), flattened to \( \latentimg_{\text{flat}} \in \mathbb{R}^{B \times N \times C} \) where \( N = H \times W \). We use multi-head attention, with \( h \) heads, each processing a subspace of dimension \( d_{\text{head}} = C / h \):
\[
Q_i = \latentimg_{\text{flat}} W_{q,i}, \quad K_i = \latentimg_{\text{flat}} W_{k,i}, \quad V_i = \latentimg_{\text{flat}} W_{v,i},
\]
where \( W_{q,i}, W_{k,i}, W_{v,i} \in \mathbb{R}^{C \times d_{\text{head}}} \). Attention scores are computed as:
\[
A_{\text{head},i} = \frac{Q_i K_i^\top}{\sqrt{d_{\text{head}}}} \in \mathbb{R}^{B \times N \times N},
\]
with outputs aggregated across heads:
\[
\text{Attention}_{\text{self}} = \text{Concat}(\text{Attention}_{\text{head},1}, \dots, \text{Attention}_{\text{head},h}) W_o,
\]
where \( W_o \in \mathbb{R}^{C \times C} \). This enables global spatial reasoning, crucial for maintaining coherence across different views in the final panoramic image.

\subsubsection{Cross-Attention}
\label{sec:cross-att}
Cross-attention integrates context embeddings \( \condctx \in \mathbb{R}^{B \times 77 \times 1024} \) with \( \latentimg_{\text{flat}} \)
\[
Q_i = \latentimg_{\text{flat}} W_{q,i}^{\text{cross}}, \quad K_i = \condctx W_{k,i}^{\text{cross}}, \quad V_i = \condctx W_{v,i}^{\text{cross}},
\]
where \( W_{q,i}^{\text{cross}} \in \mathbb{R}^{C \times d_{\text{head}}} \), and \( W_{k,i}^{\text{cross}}, W_{v,i}^{\text{cross}} \in \mathbb{R}^{1024 \times d_{\text{head}}} \). The output is
\[
\text{Attention}_{\text{cross}} = \text{Concat}(\text{Attention}_{\text{cross},1}, \dots, \text{Attention}_{\text{cross},h}) W_o^{\text{cross}},
\]
ensuring inpainted regions align with the spatial context, preserving consistent features like textures and lighting across multiple input views.

\subsection{Inpainting Model Architecture and Inputs}
\label{sec:supp-architecture}

The architecture backbone is based on the Stable Diffusion 2.1 inpainting model~\cite{stablediffusioninpaint}, which uses an encoder--decoder UNet architecture with embedded self-attention and cross-attention layers for multi-scale reasoning. The denoising process follows the DDPM framework~\cite{ho2020denoising}, where a noisy latent representation is progressively refined to reconstruct the image.

As described in the main paper,
we adapt the pre-trained inpainting diffusion model $\model(\latentimg_{\text{crop},t}^{(i)}, t, \cond )$ where
\begin{equation}
\cond = \{\mask, (1-\mask) \odot \latentimg_\text{crop}^{(i)}, \condctx \},
\end{equation}


\paragraph{Noisy Latent Input.} During training, the input noisy latent input $\latentimg^{(i)}_{\text{crop},t}$ is generated by corrupting the encoded latent of a random crop from the input panorama set $\{\imgpanos{i}\}_{i=1}^{N}$ with noise, as
\[
\latentimg_\text{crop}^{(i)} = \sdencoder(\textsc{RandCrop}(\imgpanos{i})),
\]
where the noisy latent is then
\[
\latentimg^{(i)}_{\text{crop},t} = \sqrt{\alpha_t}\latentimg^{(i)}_{\text{crop}} + \sqrt{1-\alpha_t}\noise
\]

\paragraph{Conditioned Input and Mask.}
Following Stable Diffusion 2.1, the inpainting mask $\mask \in \{0, 1\}^{B \times 1 \times 64 \times 64}$ is a downsampled representation of the region to be inpainted. It is constructed by combining random shapes with warped boundary regions to simulate occlusion patterns. The conditioning input, $\latentimg_\text{crop}^{(i)}$, is masked by $(1 - \mask)$ to simulate occlusion and then concatenated with the mask itself and the noised latent $\latentimg_t$ as input to the UNet.

\paragraph{Context Embeddings.} The embeddings $\condctx \in \mathbb{R}^{B \times 77 \times 1024}$ are derived from the same crop dimensions used for $\latentimg_\text{crop}^{(i)}$, applied to the positional encoding map $\imgpos$ and passed through the context encoder $\ctxencoder$:
\[
\condctx = \ctxencoder(\textsc{RandCrop}(\imgpos)).
\]
These embeddings replace text conditioning and are fed into the cross-attention blocks of the UNet (see Section~\ref{sec:cross-att}), enabling spatially-aware denoising.
This conditioning scheme enables the model to perform both inpainting and outpainting with spatial coherence.

\subsection{Training}
\label{sec:supp-training}

Training begins by sampling a latent representation $\latentimg_\text{crop}^{(i)}$ and the corresponding positional embedding $\condctx$ as explained in ~\ref{sec:supp-architecture}. The diffusion model $\model(\latentimg_{\text{crop},t}^{(i)}, t, \cond )$ is trained to predict the noise added to the latent during the forward diffusion process by minimizing the loss function Equation~\ref{eqn:noise}.

\paragraph{Optimization Strategy.}
The model parameters are optimized using the AdamW optimizer. To preserve the generative power of the pre-trained Stable Diffusion model, we adopt a selective fine-tuning approach:
\begin{itemize}
    \item \textbf{LoRA (Low-Rank Adaptation)}~\cite{hu2022lora} is applied to the self-attention layers of the UNet to enable efficient adaptation with fewer trainable parameters. We use a learning rate of $\num{1e-4}$.
    \item \textbf{Cross-attention layers} are fully fine-tuned to allow better integration of positional context via $\condctx$. We use a learning rate of $\num{3e-4}$.
    \item The \textbf{VAE encoder/decoder} and other layers of the UNet remain frozen to retain the fidelity of the original image reconstruction.
    \item The \textbf{context encoder $\ctxencoder$}, which produces $\condctx$, is trained from scratch using standard initialization. We use a learning rate of $\num{8e-4}$.
\end{itemize}

\subsection{Dataset}
\label{sec:supp-data}
\begin{figure*}[!htbp]
  \includegraphics[width=5in]{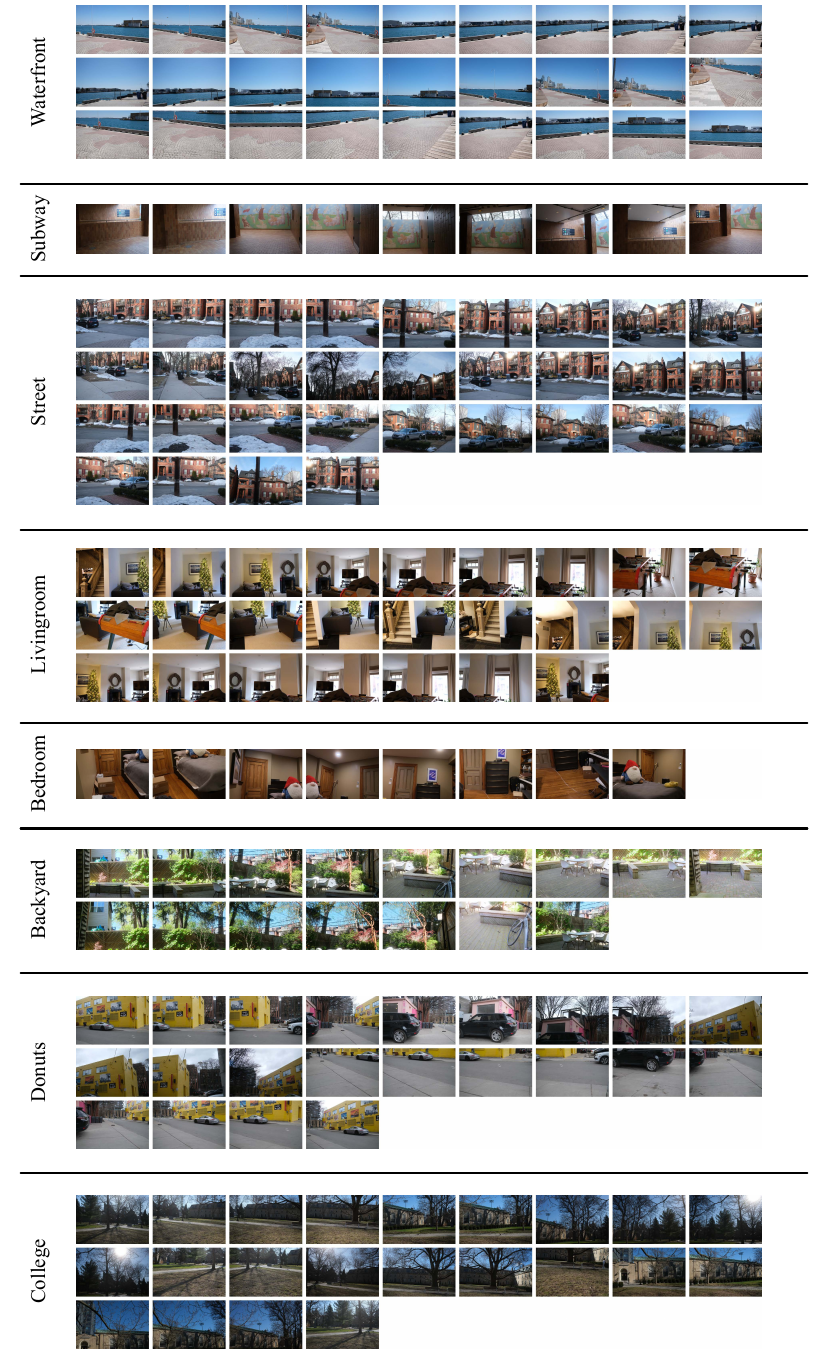}
  \caption{Reference images for the the tripod-captured dataset. Eight scenes were captured, showing a range of contexts.}
  \label{fig:tripod-dataset}
\end{figure*}

\begin{figure*}[!htbp]
  \includegraphics[width=5in]{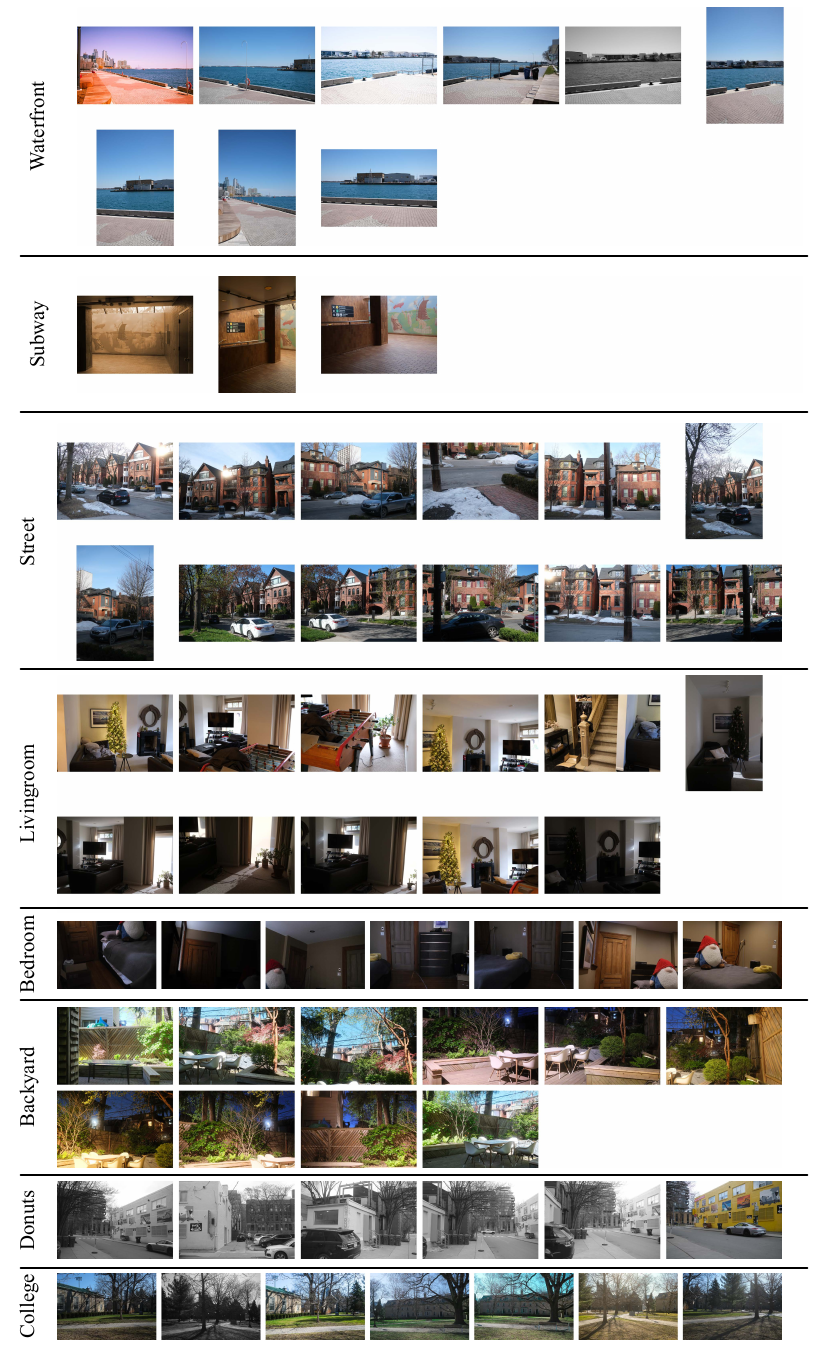}
  \caption{Reference images for the casually captured dataset. Eight scenes were captured, mirroring the tripod-captured dataset.
  This dataset shows challenging scenarios, where the input images have strong parallax effects (``Waterfront'', ``Donuts''), variations in style (``Waterfront'', ``Subway''), illumination (``Bedroom'', ``Backyard''), color palette (``Waterfront'', ``Donuts'', ``College''), or seasonal changes (``Street'').}
  \label{fig:casual-dataset}
\end{figure*}

Figure~\ref{fig:tripod-dataset} and Figure~\ref{fig:casual-dataset} show the reference images for the tripod-captured and casually-captured datasets, respectively. Both include the same eight scenes, captured (for the most part) at the exact same time: variations in lighting or time of day are captured in the casually captured set. The tripod-captured dataset attempts to capture a complete coverage of the scene, while the casually-captured dataset includes variations that make conventional stitching methods like AutoStitch~\cite{brown2007automatic} difficult. Specifically, we outline the variations for each scene as follows:
\begin{tcolorbox}[promptbox={``Backyard'' scene}]
Lighting/time-of-day (night vs. day), small parallax
\end{tcolorbox}

\begin{tcolorbox}[promptbox={``Bedroom'' scene}]
Lighting (lights on vs. off), small parallax
\end{tcolorbox}

\begin{tcolorbox}[promptbox={``College'' scene}]
White balance, image color filtering
\end{tcolorbox}

\begin{tcolorbox}[promptbox={``Donuts'' scene}]
Image color filtering, strong parallax
\end{tcolorbox}

\begin{tcolorbox}[promptbox={``Livingroom'' scene}]
Lighting variation (lights on vs. off), strong parallax, missing objects (foosball table removed), image orientation (landscape, portrait)
\end{tcolorbox}

\begin{tcolorbox}[promptbox={``Street'' scene}]
Seasonal variation (winter vs. summer), small parallax, different objects (cars), image orientation (landscape, portrait)
\end{tcolorbox}

\begin{tcolorbox}[promptbox={``Subway'' scene}]
Image color filtering (sepia), image orientation (landscape, portrait)
\end{tcolorbox}

\begin{tcolorbox}[promptbox={``Waterfront'' scene}]
Strong parallax, white balance, image color filtering, image orientation (landscape, portrait)
\end{tcolorbox}

\subsection{Baseline Implementation Details}
\label{sec:supp-baseline}
\paragraph{RealFill \cite{tang2024realfill}.} We follow the default implementation guidelines and code for RealFill. We use the default prompt ``a photo of sks'' and guidance scale of $0.99$ (as used in the original work) during inference. RealFill employs a simple prompt fine-tuning strategy as proposed in \cite{ruiz2023dreambooth}, where each reference image is randomly cropped during training and fine-tuned with the same input prompt. RealFill uses random masking during training and samples each reference image with equal probability.

\paragraph{SD2 Inpainting.} We use ChatGPT to generate the following text prompts to inpaint scenes in our dataset using the Stable Diffusion 2 inpainting \cite{stablediffusioninpaint} baseline. We fed it the reference images and asked it to describe the scene for an in/outpainting task. We use the default guidance scale of $7.5$.
\begin{tcolorbox}[promptbox={``Backyard'' scene prompt}]
A cozy backyard garden patio on a sunny spring afternoon, with light wooden fencing arranged in a chevron pattern enclosing the space. The ground is a mix of wooden decking and brick pavers. There is a modern white outdoor dining table surrounded by white molded chairs with wooden legs. Raised garden beds line the perimeter, filled with lush green ferns, hostas, and flowering tulips. A small Japanese maple tree with red leaves adds a vibrant accent. Overhead, string lights hang between tall trees. Suburban townhouses and fire escapes are visible beyond the fence. Continue the garden with more raised beds, greenery, and cozy shaded seating areas.
\end{tcolorbox}

\begin{tcolorbox}[promptbox={``Bedroom'' scene prompt}]
A cozy bedroom with warm lighting and natural wood trim. A soft grey bedspread covers a modern dark wood bed, with a large plush gnome toy sitting upright at the head of the bed. The gnome has a fluffy white beard, red hat, and blue outfit. To the side, there’s a pair of bright yellow slippers on the bed. The walls are painted beige, with framed artwork and a tall wooden door. Hardwood floors reflect the warm overhead light. The room is neat but lived-in, with a cardboard box on the floor, a dresser topped with a camera and board games, and open shelves filled with books and gadgets. Extend the scene naturally with matching lighting, wood finishes, and layout.
\end{tcolorbox}

\begin{tcolorbox}[promptbox={``College'' scene prompt}]
A serene university courtyard on a crisp early spring morning, lined with tall leafless trees casting long shadows across the stone-paved paths. Old gothic stone buildings and vintage street lamps border the green lawn, while scattered wooden benches sit empty under the bare branches. A soft blue sky with gentle morning sunlight peeks through the trees, illuminating the historic campus in a calm, peaceful atmosphere. Natural lighting, detailed textures, realistic architecture, high-resolution photo.
\end{tcolorbox}

\begin{tcolorbox}[promptbox={``Donuts'' scene prompt}]
Urban back alley beside a bright yellow building with murals depicting industrial and artistic scenes, a red ‘RECEIVING’ sign, and barred windows. A silver sports car is parked on the street beside a city parking meter. Adjacent to it is a pastel pink storefront with bold ‘HOT DOG’ signage, garbage bins, and leafless winter trees. In the distance, high-rise glass apartments and brick institutional buildings frame the cityscape. Overcast sky, soft urban lighting, clean and calm street scene.
\end{tcolorbox}

\begin{tcolorbox}[promptbox={``Livingroom'' scene prompt}]
Extend the cozy living room with warm lighting and rustic cabinetry, holiday decor continuing throughout the space.
\end{tcolorbox}

\begin{tcolorbox}[promptbox={``Street'' scene prompt}]
A quiet urban residential street in winter, lined with large Victorian red-brick houses with gabled roofs and stone foundations. Bare trees stand on small front lawns covered in patches of snow. Parked cars line the street, including compact sedans and hatchbacks. The sky is clear and blue, with soft late afternoon sunlight casting long shadows. A mix of brick textures, wooden porches, and balconies add architectural charm. Extend the street with similar architecture, snow-covered sidewalks, more houses in perspective, and consistent lighting and color tone.
\end{tcolorbox}

\begin{tcolorbox}[promptbox={``Subway'' scene prompt}]
A vintage subway concourse with brown ceramic tiles and steel railings, illuminated by ceiling spotlights. A surreal mural on the wall shows whimsical floating objects, vintage figures, a red car, a lion on a bed, and abstract staircases, all set against a bright grassy hill and a blue sky with clouds. Large glass windows above let in natural daylight and reveal leafless tree branches outside. The overall atmosphere is clean, quiet, and dreamlike, blending realism with surrealism.
\end{tcolorbox}

\begin{tcolorbox}[promptbox={``Waterfront'' scene prompt}]
Toronto waterfront promenade on a clear sunny day, with a patterned brick walkway, stone barriers, life buoys on silver poles, modern industrial dock buildings across the lake, calm blue water, Porter Airlines ferry terminal, and distant city skyline with high-rise towers, boats on the lake, realistic urban scenery, vibrant shadows and natural lighting
\end{tcolorbox}

\subsection{Metrics}
\label{sec:supp-metrics}

To evaluate the quality of generated panoramic images, we employ a comprehensive set of metrics that assess standard image quality, high-level image structure, and preservation of scene layout. The latter two use learning-based metrics and feature-matching-based metrics, respectively. We evaluate generated panoramas \( \imgpano^{\text{gen}} \in \mathbb{R}_{+}^{H_\text{pano} \times W_\text{pano} \times 3} \) against the reference panorama produced using AutoStitch~\cite{brown2007automatic} on the tripod-captured dataset \( \imgpano^{\text{ref}} \in \mathbb{R}_{+}^{H_\text{pano} \times W_\text{pano} \times 3} \). All metrics except DreamSim, CLIP, and DINO omit the region of the sparse panorama provided as input during inference (defined by a binary mask \( \mask_{\text{input}} \in \{0, 1\}^{H_\text{pano} \times W_\text{pano}} \)) to focus on inpainted areas. Below, we describe each metric, its implementation details, and its significance.

\paragraph{PSNR (Peak Signal-to-Noise Ratio).} Measures pixel-level similarity~\cite{hore2010image}, defined as
\[
\texttt{PSNR} = 10 \cdot \log_{10} \left( \frac{\texttt{MAX}_{\bf{x}}^2}{\texttt{MSE}(\imgpano^{\text{ref}}, \imgpano^{\text{gen}})} \right),
\]
where \(\texttt{MAX}_{\bf{x}} = 255\) for 8-bit RGB images, and \(\texttt{MSE}(\cdot)\) is the mean squared error between valid pixels (i.e., where \( \mask_{\text{input}} = 0 \)) of \( \imgpano^{\text{ref}} \) and \( \imgpano^{\text{gen}} \). Higher values indicate better pixel fidelity.

\paragraph{SSIM (Structural Similarity Index).} Assesses structural and perceptual similarity by comparing luminance, contrast, and structure in grayscale images~\cite{wang2004image}:
\[
\texttt{SSIM}(x, y) = \frac{(2\mu_x\mu_y + c_1)(2\sigma_{xy} + c_2)}{(\mu_x^2 + \mu_y^2 + c_1)(\sigma_x^2 + \sigma_y^2 + c_2)},
\]
where \(\mu_x, \mu_y\) are means, \(\sigma_x, \sigma_y\) are variances, \(\sigma_{xy}\) is covariance, and \(c_1, c_2\) are constants. We compute SSIM on grayscale images with masked regions (\( \mask_{\text{input}} = 1 \)) set to zero. Higher values indicate better structural consistency.

\paragraph{LPIPS (Learned Perceptual Image Patch Similarity).} Measures perceptual similarity using a pre-trained AlexNet~\cite{krizhevsky2012imagenet}, as proposed by Zhang et al.~\cite{zhang2018unreasonable}. For permuted image tensors \( \imgpano^{\text{ref}}, \imgpano^{\text{gen}} \in \mathbb{R}_{+}^{B \times 3 \times H_\text{pano} \times W_\text{pano}} \), normalized to \([-1, 1]\), LPIPS computes feature distances as
\[
\texttt{LPIPS} = \frac{1}{H'W'} \sum_{h,w} \texttt{loss}_{\texttt{Alex}}(\imgpano^{\text{ref}}, \imgpano^{\text{gen}}) \cdot (1 - \mask_{\text{input}}'),
\]
where \(\texttt{loss}_{\texttt{Alex}}\) is the weighted L2 distance between feature activations from AlexNet layers, \( \mask_{\text{input}}' \) is the resized mask, and \( H', W' \) match the feature map size. Lower values indicate better perceptual similarity.

\paragraph{DreamSim~\cite{fu2023dreamsim}.} Evaluates high-level perceptual similarity using the DreamSim model trained on human perceptual judgments. The metric is:
\[
\texttt{DreamSim} = \texttt{dreamsim\_model}(\imgpano^{\text{ref}'}, \imgpano^{\text{gen}'}),
\]
where \( \imgpano^{\text{ref}'}, \imgpano^{\text{gen}'} \) are images resized to \(224 \times 224\) to match the model’s input requirements. The DreamSim model, based on a vision transformer, predicts perceptual similarity by comparing feature embeddings. Lower scores indicate closer perceptual alignment.

\paragraph{DINO.} Measures semantic similarity using the DINOv2-base model~\cite{oquab2023dinov2} as the cosine distance between features extracted from the last hidden state:
\[
\texttt{DINO} = \frac{\cos(\texttt{feat}^{\text{ref}}, \texttt{feat}^{\text{gen}}) + 1}{2},
\]
where \(\texttt{feat}^{\text{ref}}, \texttt{feat}^{\text{gen}}\) are mean-pooled features from the last hidden state of DINOv2, and \(\cos(\cdot)\) is the cosine similarity. Higher values indicate better semantic alignment.

\paragraph{CLIP.} Assesses semantic similarity using the CLIP ViT-B/32 model~\cite{radford2021learning}. The CLIP score is defined as the cosine similarity between normalized CLIP image embeddings:
\[
\texttt{CLIP} = \cos(\latentimg^{\text{ref}}_{\texttt{clip}}, \latentimg^{\text{gen}}_{\texttt{clip}}),
\]
where \(\latentimg^{\text{ref}}_{\texttt{clip}}, \latentimg^{\text{gen}}_{\texttt{clip}}\) are image embeddings from the forward pass of the CLIP ViT-B/32 model. Higher values indicate better semantic consistency.


\begin{figure}[h]
  \centering
  \includegraphics[width=3.5in]{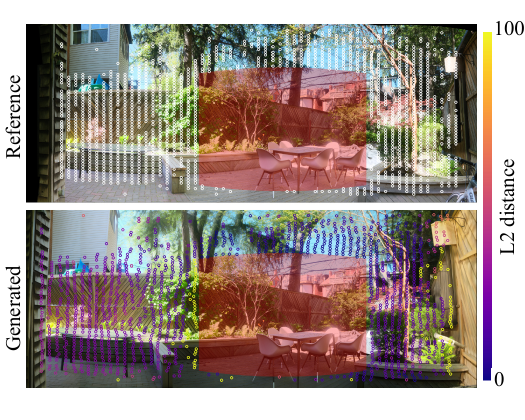}
  \vspace{-1em}
  \caption{Example of LoFTR feature matching between reference and generated panoramas. White circles mark selected keypoints in the reference, with corresponding points in the generated panorama color-coded by L2 pixel distance from their reference positions.}
  \label{fig:loftr}
\end{figure}

\paragraph{LoFTR Metrics.} Evaluates feature correspondence using LoFTR~\cite{sun2021loftr}. Images are resized to \( 512 \times 512 \) and converted to grayscale and then processed. We report both the L2 distance between the pixel coordinates of matching features and the number of matched features divided by the total number of features in the reference image. Specifically,
\begin{align}
&\texttt{LoFTR\_L2\_Distance} = \frac{1}{N} \sum_{i=1}^N \sqrt{\sum (\texttt{mkpts}^{\text{ref}}_i - \texttt{mkpts}^{\text{gen}}_i)^2}, \nonumber \\
&\texttt{LoFTR\_Match\_Proportion} = \frac{N}{\texttt{total\_features}}, \nonumber
\end{align}
where \(\texttt{mkpts}^{\text{ref}}, \texttt{mkpts}^{\text{gen}}\) are matched keypoints outside \( \mask_{\text{input}} \), \( N \) is the number of valid matches between the reference and the generated panorama identified by LoFTR, and \(\texttt{total\_features}\) is the number of reference keypoints identified by LoFTR in the reference panorama. Lower \(\texttt{LoFTR\_L2\_Distance}\) and higher \(\texttt{LoFTR\_Match\_Proportion}\) indicate better correspondence. An example of matches identified by LoFTR for an image crop is visualized in Figure~\ref{fig:loftr}.

\subsection{Inference Procedure Details}
\begin{figure*}[h]
  \includegraphics[width=5.52in]{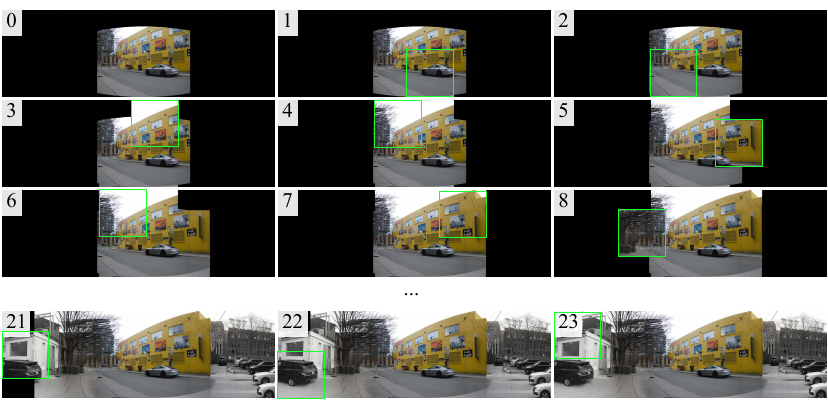}
  \caption{Example of the tiling strategy to generate the full panoramic canvas. Tiles closest to the reference image are denoised first, with subsequent tiles denoised in a breadth-first-search manner.}
  \label{fig:tiling}
\end{figure*}

During inference, we adopt a tile-based approach to progressively generate the full panoramic canvas. For each tile, the model performs \( T \) denoising steps, leveraging the concatenated input and cross-attention-guided context embeddings \( \condctx \) to produce the tile output. We used $T=50$. This procedure is repeated for all tiles, and the final panoramic image \( \imgpano^{\text{gen}} \) is assembled sequentially, tile by tile. Tiles are sorted by the distance to the centroid of the starting reference image, in increasing distance, in a breadth-first-search manner. This allows the denoising of tiles with overlap of the starting reference image first, and subsequently outpainting tiles with overlap from previous generations. An example is shown in Figure~\ref{fig:tiling}.

\begin{figure*}[h]
  \includegraphics[width=5.52in]{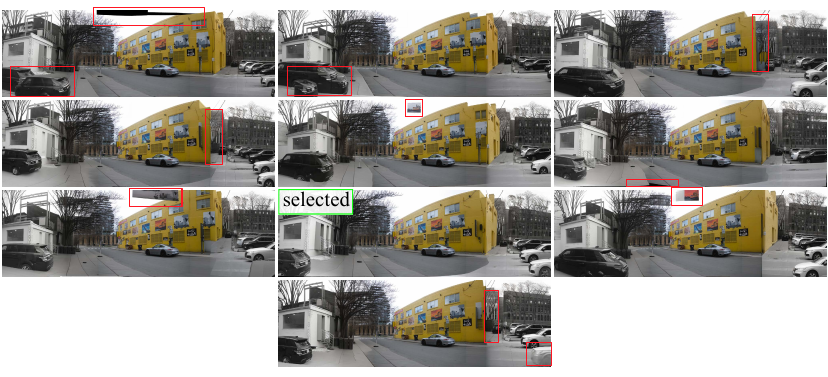}
  \caption{Example of the correspondence-based seed selection strategy to generate the full panoramic canvas. We generate 10 panoramas with different seeds and select the one with most feature matches. The selected panorama has the least artifacts in this example and is the most seamless and similar to the reference.}
  \label{fig:seeds}
\end{figure*}

\paragraph{Correspondence-based seed selection.} 
Due to stochasticity in the inference process, the generation quality varies between random seeds. This is amplified by the numerous tiles required to denoise a full panorama, and artifacts early-on may propagate throughout the canvas.  
We employ a correspondence-based seed selection process~\cite{tang2024realfill} to mitigate this problem, identifying generated panoramas whose layout matches the result of feature-based image registration~\cite{brown2007automatic}. An example of various seed generations is shown in Figure~\ref{fig:seeds}.
We generate ten panoramas with different random seeds and take our output to be the panorama with the most feature matches (computed with LoFTR~\cite{sun2021loftr}) compared to the output of AutoStich \cite{brown2007automatic} on the casually-captured dataset. Final metrics would be calculated by comparing the reference panorama from the tripod-captured dataset. This process could be further enhanced with more seeds, depending on desired computation budget (e.g. RealFill~\cite{tang2024realfill} generates 64 outputs).

\section{Supplementary Results}
\label{sec:supp-results}

\subsection{Tripod-Captured Dataset}
\begin{figure*}[!htbp]
  \includegraphics[width=5.52in]{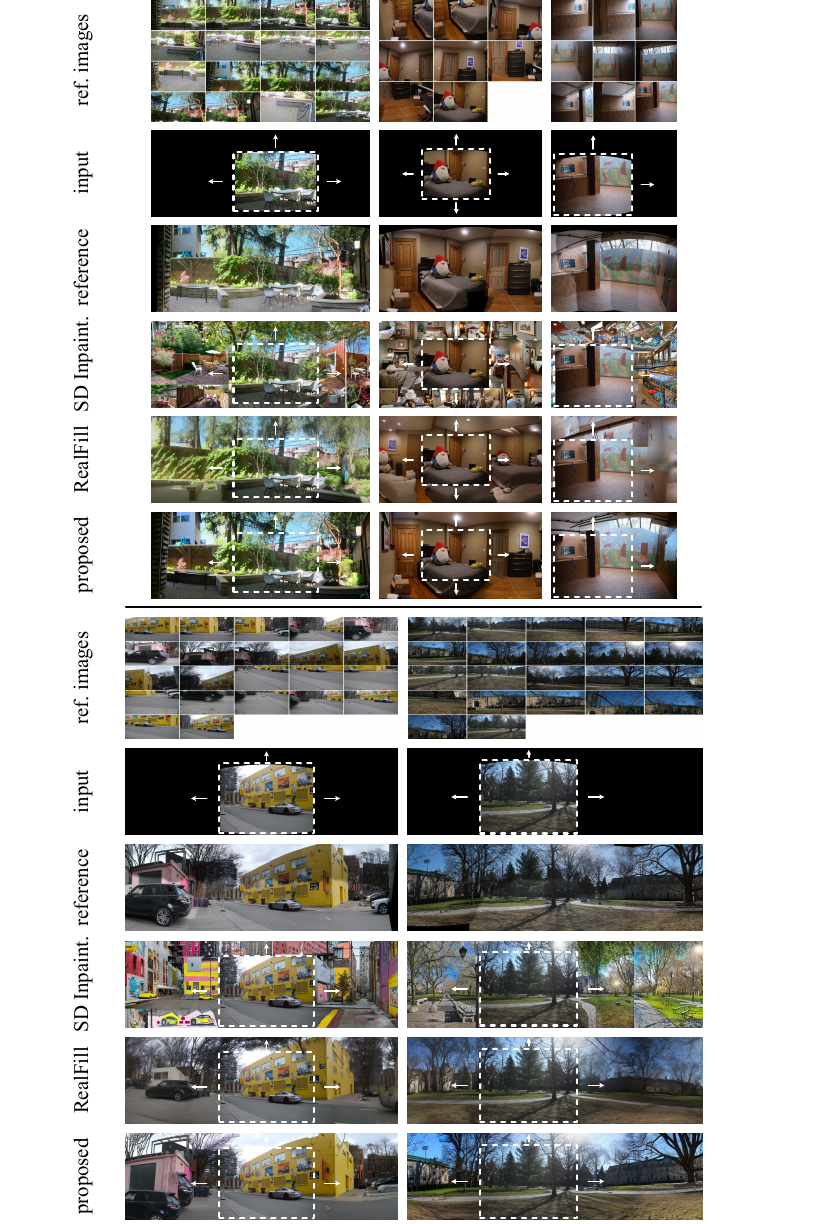}
  \caption{Qualitative results on the additional scenes from the tripod-captured dataset. We find that our approach produces panoramas that are more consistent with the layout and content of the reference panorama than baseline approaches based on inpainting/outpainting.}
  \label{fig:supp-tripod}
\end{figure*}

We show the additional 5 scenes for the tripod-captured dataset in Figure~\ref{fig:supp-tripod}.
Similar to before, we observe
that the Stable Diffusion inpainting model~\cite{stablediffusioninpaint} produces image content that is locally plausible, but
fails to adhere to the layout and content of the actual scene. Similar to previous scenes, RealFill~\cite{tang2024realfill} improves on this result, but tends to repeat scene content and ignores scene layout.
Our approach provides a much closer match to the layout provided by the reference panorama.

\subsection{Casually Captured Dataset}
\begin{figure*}[!htbp]
  \includegraphics[width=5.52in]{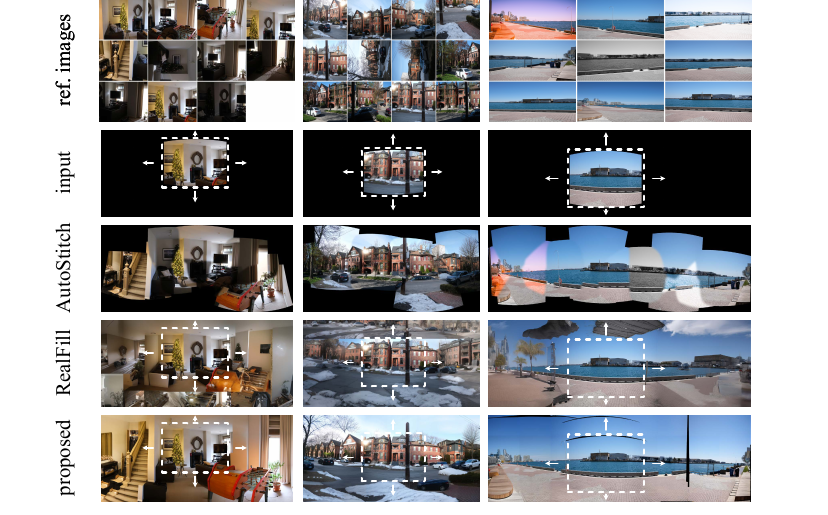}
  \caption{Qualitative results of the additional three scenes on the casually captured dataset. Even in this challenging scenario, where the input images have strong parallax effects and variations in style, illumination, color palette, or camera capture settings, our approach reconstructs seamless panoramas that preserve the content and layout of the reference.}
  \label{fig:supp-casual}
\end{figure*}

We show the additional three scenes for the casually captured dataset in Figure~\ref{fig:supp-casual}.
Similar to other scenes, AutoStitch~\cite{brown2007automatic}, fails to convincingly
blend between the different image regions, resulting in ghosting and other artifacts.
RealFill exhibits similar artifacts as in the tripod-captured dataset, and we find that our approach produces
seamless results that are more consistent with the layout and content of the scene. 

\subsection{Supplementary Ablation Studies}
\label{sec:supp-ablations}



\paragraph{Quantitative results.}
We conduct an ablation study on the casually captured dataset (see  Table~\ref{tab:supp-results}). We evaluate (1) the effects of parameter choices in positional encoding frequencies (number of channels, max frequency, and omitting token positional encodings), (2) inference strategies, omitting the reference image during inference and denoising tiles row-by-row, with rows sorted by distance to the starting image in the y-axis, and tiles sorted by the distance to the centroid in the x-axis, (3) various guidance scales, (4) various overlap ratios, and (5) training without positional encoding and only using warped reference images. Each ablation uses correspondence-based seed selection to eliminate concerns over seed selection.

Significantly lower max frequency (10Hz), smaller number of channels (4 channels), and no token positional encoding show improvements in some image quality metrics, but a fall in the feature-matching-based metrics. The higher frequencies of the proposed method (12-channels, 50Hz) allow for finer details and better reconstruction of features from the reference images.

Removing the reference image shows a drop across the board in performance, showing the necessity of a starting reference image, as expected. Performance still outperforms prior baselines (see Table~\ref{tab:casual}).


Guidance scales between $1.5$ and $2.00$ and overlap ratios between $0.1$ and $0.2$ show the best performance in class. We chose a guidance scale of $1.5$ with an overlap ratio of $0.2$. Generating panoramas using a row-by-row sorting shows marginal improvements in some metrics, however, we found qualitatively that more artifacts are produced. These artifacts are more evident to users, and therefore we opted not to use this strategy.

\begin{wrapfigure}[13]{r}{3in}
\vspace{-1.5em}
\includegraphics{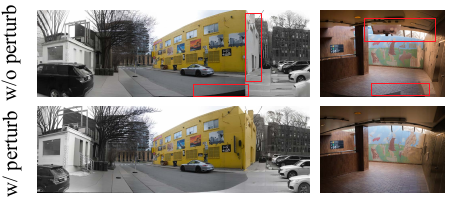}
\caption{Qualitative evaluation of the ablation omitting the similarity transform used to perturb the location of the warped images in the sparse panoramas.}
  \label{fig:pert}
\end{wrapfigure}

RealFill with warped reference images suffers from similar repetitive content and a lack of adhesion to the reference layout, demonstrating the need for our proposed positional encoding conditioning.

\begin{table}[!tbp]
    \small
    \centering
    \setlength{\tabcolsep}{1pt}
    \resizebox{\textwidth}{!}{%
    \begin{tabular}{lcccccccc}
         \toprule
         Method &  PSNR $\uparrow$ & SSIM  $\uparrow$& LPIPS $\downarrow$ & DreamSim $\downarrow$ & DINO $\uparrow$ & CLIP $\uparrow$ & LoFTR (L2 Distance) $\downarrow$ & LoFTR (Matching) $\uparrow$\\ \toprule
         proposed (10Hz)              &                      11.09 (2.03) & \cellcolor{tabsecond}0.414 (0.100) &                      0.532 (0.065) &                      0.137 (0.024) &                      0.971 (0.011) &                      0.917 (0.034) &                      21.65 (4.48) &                      0.119 (0.037) \\
proposed (4 channels)        &                      11.12 (1.96) &  \cellcolor{tabfirst}0.416 (0.102) &                      0.543 (0.065) & \cellcolor{tabsecond}0.134 (0.013) &  \cellcolor{tabthird}0.973 (0.011) &                      0.914 (0.042) &                      21.95 (5.54) &                      0.109 (0.039) \\
proposed (w/o token pos enc) &                      10.99 (2.29) &  \cellcolor{tabthird}0.413 (0.104) &                      0.544 (0.068) &                      0.149 (0.033) &  \cellcolor{tabfirst}0.975 (0.009) &                      0.913 (0.037) &                      19.89 (3.82) &                      0.114 (0.044) \\
proposed (no ref)            &                      10.20 (2.04) &                      0.311 (0.138) &                      0.666 (0.067) &                      0.235 (0.065) &                      0.943 (0.011) &                      0.866 (0.047) &                      26.89 (6.25) &                      0.107 (0.042) \\
proposed (guidance=0.99)     &  \cellcolor{tabfirst}12.22 (1.92) &                      0.395 (0.140) &                      0.508 (0.061) &                      0.154 (0.032) &                      0.972 (0.011) &  \cellcolor{tabthird}0.922 (0.043) &                      18.13 (4.57) &                      0.118 (0.057) \\
proposed (guidance=1.00)     &  \cellcolor{tabfirst}12.22 (1.92) &                      0.395 (0.140) &                      0.508 (0.061) &                      0.154 (0.032) &                      0.972 (0.011) &  \cellcolor{tabthird}0.922 (0.043) &                      18.13 (4.57) &                      0.118 (0.057) \\
proposed (guidance=2.00)     &                      11.07 (2.25) &                      0.363 (0.135) &                      0.514 (0.073) &                      0.145 (0.044) &                      0.972 (0.012) &  \cellcolor{tabthird}0.922 (0.038) &  \cellcolor{tabthird}17.20 (4.32) &  \cellcolor{tabfirst}0.131 (0.056) \\
proposed (guidance=3.00)     &                      10.43 (2.28) &                      0.348 (0.129) &                      0.539 (0.073) &                      0.165 (0.042) &                      0.968 (0.013) &                      0.919 (0.032) & \cellcolor{tabsecond}16.91 (5.64) &                      0.122 (0.058) \\
proposed (guidance=5.00)     &                      9.57 (2.14) &                      0.330 (0.125) &                      0.593 (0.066) &                      0.188 (0.041) &                      0.962 (0.018) &                      0.917 (0.035) &                      17.79 (6.15) &                      0.109 (0.055) \\
proposed (guidance=7.50)     &                      8.80 (1.50) &                      0.310 (0.107) &                      0.648 (0.049) &                      0.251 (0.065) &                      0.940 (0.034) &                      0.897 (0.040) &                      24.03 (15.30) &                      0.097 (0.056) \\
proposed (overlap=0.00)      &                      11.37 (2.18) &                      0.373 (0.144) &                      0.514 (0.079) &                      0.136 (0.031) & \cellcolor{tabsecond}0.974 (0.009) &                      0.919 (0.035) &                      17.82 (5.25) &                      0.124 (0.052) \\
proposed (overlap=0.10)      &                      11.34 (1.89) &                      0.372 (0.137) &  \cellcolor{tabthird}0.507 (0.077) &                      0.140 (0.033) &                      0.970 (0.012) &                      0.920 (0.030) &  \cellcolor{tabfirst}16.52 (5.02) &                      0.128 (0.053) \\
proposed (overlap=0.50)      &  \cellcolor{tabthird}11.56 (2.11) &                      0.377 (0.140) & \cellcolor{tabsecond}0.501 (0.082) &  \cellcolor{tabthird}0.135 (0.040) &                      0.970 (0.010) & \cellcolor{tabsecond}0.927 (0.029) &                      19.20 (7.96) &                      0.128 (0.055) \\
proposed (overlap=0.75)      & \cellcolor{tabsecond}11.71 (2.06) &                      0.378 (0.133) &  \cellcolor{tabfirst}0.499 (0.084) &  \cellcolor{tabthird}0.135 (0.038) &                      0.971 (0.013) &  \cellcolor{tabfirst}0.932 (0.022) &                      18.43 (9.01) &                      0.119 (0.062) \\
proposed (row-by-row)        &                      11.54 (2.16) &                      0.379 (0.142) &  \cellcolor{tabthird}0.507 (0.070) &  \cellcolor{tabfirst}0.131 (0.026) & \cellcolor{tabsecond}0.974 (0.011) &                      0.911 (0.045) &                      17.75 (4.98) &  \cellcolor{tabthird}0.129 (0.051) \\
RealFill (warped ref)        &                      10.39 (1.64) &                      0.309 (0.132) &                      0.667 (0.080) &                      0.248 (0.040) &                      0.932 (0.020) &                      0.884 (0.024) &                      60.56 (26.77) &                      0.016 (0.002) \\
proposed                     &                      11.35 (2.15) &                      0.374 (0.143) &                      0.508 (0.076) &                      0.137 (0.033) &                      0.971 (0.013) &                      0.917 (0.035) &                      17.97 (5.14) & \cellcolor{tabsecond}0.130 (0.056)
\\
         \bottomrule
 \end{tabular}}
    \captionof{table}{Ablation study. We evaluate the effects of (1) using a max frequency of 10Hz for the positional encoding (10Hz), (2) using 4 channels for the positional encoding (4 channels), (3) omitting the token positional encoding in the positional encoding (w/o token pos enc), (4) omitting the reference image during inference (no ref), (5) various guidance scales (guidance), (6) various overlap ratios (overlap), (7) tiling strategy, denoising row-by-row (row-by-row), and (8) RealFill trained with the warped reference images.
    We compare the generated results to a reference panorama produced using AutoStitch~\cite{brown2007automatic} on the tripod-captured dataset as in the main paper.}
    \label{tab:supp-results}
\end{table}

\paragraph{Qualitative results.}
We show qualitative results for the effect of perturbing the locations of sparse images in the panorama Figure~\ref{fig:pert}.
Without perturbation, the model is less robust to misalignments in the initial layout estimation.

\begin{figure*}[!htbp]
  \includegraphics[width=5.52in]{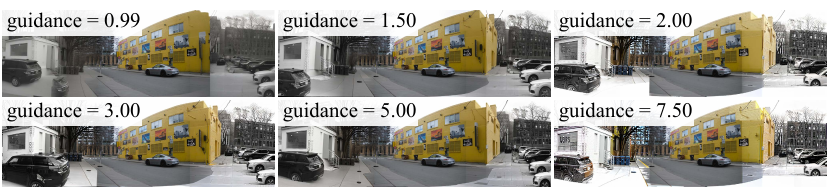}
  \caption{Qualitative comparison of various guidance scales on the casually-captured dataset. We find that increasing guidance leads to more ``cartoonish'' outputs and more seams, with a guidance scale around $1.5$ showing best results.}
  \label{fig:guid}
\end{figure*}

We show qualitative results for the various guidance scales in Figure~\ref{fig:guid}. Lower guidance scales ($<1.5$) maintain scene quality but fail to properly blend through artifacts (e.g. building remains grayscale). Guidance scales between $1.5-2$ show how scene cohesion can be maintained while also resolving artifacts found in the reference images (building is well blended). Higher guidances begin to exhibit ``cartoonish'' effects and the scene loses cohesion (obvious seams between regions in the panorama).

\end{document}